# Drawing Science: an interdisciplinary, playful, and inclusive proposal

M. A. M. Souza[a][1], V. Dutra[a], T. B. Menezes[a], J. W. Silva[b]

[a]Instituto Federal do Piauí, Parnaíba, Piauí, Brazil

[b]Universidade Federal do Ceará, Fortaleza, Ceará, Brazil

**Abstract**

This article proposes and discusses qualitatively the use of a didactic board game in Science Education for high school students. The game contemplates in an interdisciplinary way the areas of Physics, Biology, Chemistry and Astronomy and is based on the development of the cognitive structure of the students through the act of drawing the concepts addressed in the classroom, also aiming at socialization through teamwork and the inclusion of students with hearing impairment. The game was created in a physics teacher education course and also aims to promote reflection on teaching practice and the search for attractive teaching methodologies for students.

**Keywords:** Science Education, interdisciplinary, act of drawing; inclusion, teacher education.

## 1. Introduction

Many high school students adopt a posture of disinterest or even disgust towards science subjects, especially physics. It can be said that this is a problem constantly faced by teachers of Basic Education, where many of them claim that the solution lies in the establishment of adequate conditions for the development of students' cognitive skills (Ricardo, 2010; Camargo & Nardi, 2004).

With the new High School and the regulation of the Common National Curricular Base (BNCC) it was established that in High School the Natural Sciences and their technologies are integrated through the development of projects that address contemporary cross-cutting themes with the objective of promoting interdisciplinarity between the various areas of

---

[1] e-mail: msouza@ifpi.edu.br

orcid: http://orcid.org/0000-0002-0224-272X

science, such as Physics, Chemistry, Biology, and other scientific areas, such as Astronomy, Engineering etc.

It is worth mentioning here the guidelines of the Ministry of Education (MEC) for the development of the competences proposed by the National Curriculum Parameters (PCN) for the teaching of Physics and Sciences that must follow: representation and communication; research, understanding and socio-cultural contextualization. The traditional areas of Physics, Mechanics, Thermology, Optics and Electromagnetism are grouped into six themes that can be seen in Table 1.

| Theme of National Curriculum Parameters | Areas of physics |
|---|---|
| Movements: variations and conservation | Mechanics |
| Heat, environment, and energy uses | Mechanics, Thermology, Optics and Electromagnetism |
| Sound, image, and information | Mechanics, Optics and Electromagnetism |
| Electrical and telecommunications equipment | Optics and Electromagnetism |
| Matter and radiation | Thermology, Optics and Electromagnetism |
| Universe, Earth, and life | Mechanics, Thermology, Optics and Electromagnetism |
| Total | 6 themes |

Table 1. PCN themes and their correlation with the areas of physics.

The integrative projects in the new High School become an interesting tool from the moment that they provide meaningful learning for students, as the latter starts to see the meaning, functionality, and usefulness of scientific knowledge for society through the development of new technologies, the conservation of the environment, the use of renewable energy and space exploration. This is a way of promoting scientific literacy among students, as science must be disseminated among young people with the aim of awakening them to the reality of the technological society, serving as a vehicle for inclusion and as a way of stimulating the formation of new scientists.

In this new context, we created a teaching and assessment proposal using a didactic game whose main focus is to promote interdisciplinarity between Natural Sciences and the inclusion of students with special needs. From a methodological and didactic point of view, the objective of the game, is to complement interdisciplinary learning among the Natural Sciences, and it can also be used as an assessment tool, since participants must draw elements that are related to your knowledge and scientific experience, while advancing on the board. Each drawing represents the student's cognitive structure and refers to his or her worldview and will be subject to interpretation and debate among the other students who are part of the game group. With gamification (or ludification) in teaching, we seek to develop cognitive ability (Almeida et al, 2016) and students' concentration, the interpretative and investigative sense, because every move can lead to a new discovery, a new concept, in addition to having scientific collaboration within the groups of players as a goal. A positive experience with the use of educational games for teaching particle physics can be found at (Souza et al, 2019).

From this perspective, in the next subsections we will describe general aspects of teacher education in Brazil, we will address aspects of inclusion in the education of deaf people in Brazilian schools and we will discuss the relationship between the act of drawing and meaningful learning. In section 2 we will present the methodology used and the rules of the game, then in section 3, we will present the results and discuss the potentials, limitations and applicability of the game and the final considerations will be made in section 4.

1.1  Teacher Education

The education of teachers for Basic Education is a delicate topic and needs to be taken care of in our society. Historically it started in 17th century France, with the Masters' seminar, created by Saint John the Baptist of La Salle, aiming at the formation of teachers to teach poor children, in a religious congregation, the Brothers of the Christian Schools, or Lasallian Brothers. But it was only in the 19th century, with the inspiration of the ideals of the French Revolution, that the creation of regular schools for teacher education was officially instituted. This process only started in Brazil with Independence and later with the advent of the proclamation of the Republic, having gained strength in the 1950s with the country's industrialization process, which also led to the creation of Technical schools (Tanuri, 2000; Saviane, 2008; Saviane, 2009).

Between the 40s and the 70s, Pedagogy and Licentiate courses were implemented in Brazil, however the vision used to implement these courses and the methodologies used based on cultural-cognitive models, in which teacher education is based only in the domain of content which will be taught, led to some problems that are still present today in Higher Education Institutions (Saviane, 2007). There is a mistaken view that the teacher does not need pedagogical preparation, it is enough to know in general the knowledge that will be passed on, not giving due importance to the teaching methodology, that is, how to teach and why to teach. In addition to this historical vision of formation, we still have serious structural problems in our public schools, with scrapped or nonexistent classrooms and laboratories, lack of incentive and devaluation of the teaching career.

In many cases, physics teachers are being trained based on a model that prioritizes the transmission of knowledge without proper contextualization (Pietrocola et al, 2003), without establishing a relationship with scientific and technological development, most of the time there is an excessive routine of expository classes and resolution of exercises that, in general, prioritize the memorization of mathematical formulas. It is necessary to seek methodologies that arouse students' interest and fascination for science.

1.2  Inclusive education for the deaf in Brazilian schools

The term Inclusive Education refers to equal rights and conditions of access for all students in institutions of regular education, at all levels of education (Hammeken, 2007; Armstrong et al, 2010; Hornby, 2014; Reichow et al, 2016). Inclusive Education advocates that all students should be together, sharing and enjoying the same educational resources and structures, without any type of discrimination, bullying and harassment. It is the defense of diversity aiming to meet the educational individualities of each subject in the classroom to promote the cognitive development of all, regardless of race, gender, financial situation, religion and physical or mental disability.

In Brazil, assistance to people with disabilities began in the imperial period, when two institutions were created to serve special students: the Imperial Institute for Blind Children, founded in 1854, which today is known as Benjamin Constant Institute (IBC); and the Deaf Mute Institute, founded in 1857, which today is known as the National Institute for the Education of the Deaf (INES), both in the city of Rio de Janeiro (Mazzotta, 2017; Jannuzzi, 2012).

In 1996, the Law of Guidelines and Bases of Education (LDB) recognized the need for inclusion and respect for the individuality of each student, establishing in its article 58 the creation of Special Education as a modality of school education, offered to students with special needs (LDB nº 9.394/96).

About the education of the deaf in Brazil, 2002 was marked by the regulation of the Brazilian Sign Language (Libras) through Law 10.436, which also established the need for the presence of a translator or interpreter in educational institutions. Subsequently in 2005, Law 5.626 was enacted, instituting the teaching of Libras for all teacher training courses, be they at higher, secondary, or elementary levels.

The main difficulty in teaching deaf students in Brazilian schools is precisely due to communication, because nowadays, even with the discipline of Libras in undergraduate courses, the teacher is not yet fully prepared to meet the special needs, in the classroom, of a deaf student. Another problem is that most schools do not have an interpreter, which makes accessibility even more difficult.

Teaching science to deaf students is a real challenge, it is of utmost importance to choose a methodology that can provide the student with a clear understanding of the concept to be explained. A very attractive way that can generate good results in the teaching-learning process is education through a visual language, that is, the educator can use the so-called imagery language (Campello, 2008). Such language includes the use of various resources, which can be of a gestural nature, using the body itself, screens, walls, school notebooks, drawings, pictures, and the whiteboard in the classroom. And it is at this point that the game Drawing Science, which we are proposing fits within the perspective of Inclusive Education for deaf students, as it uses body gestures and the act of drawing on the whiteboard in the classroom to assign meaning to a scientific concept.

1.3  The act of drawing and meaningful learning

In meaningful learning, the student develops his cognitive structure through the interaction of new knowledge with his previous knowledge of a given subject. It is anchored in the idea of the existence of subsunitors, that is, constructs that integrate information with the human brain, function as an anchor for the new learned concept, which will be stored hierarchically and organized in the human mind (Ausubel, 2000; Ausubel et al, 1980).

The use of diagrams, drawings, and the act of drawing in the classroom, perfectly permits the mapping of the student's cognitive development, further contributing to constructivist learning (Webb & Hassen, 1988; Paivio, 2006; Souza & Duarte, 2015; McLure et al, 2021). The creation of drawings by the students facilitated the construction of relationships and meanings, favoring the meaningful learning of the students, also helping to:

1) Raise the level of the cognitive process, requiring students to think at a higher level, generalizing concepts and relationships.
2) Require students to define their ideas more precisely.
3) Differentiate general knowledge from specific knowledge.
4) Improve the analytical and intuitive skills, leading to a systematization and organization of the acquired knowledge.
5) Provide opportunities for students to test their own cognitive models, detect and correct inconsistencies.

Drawing is the oldest form of human language, dating back to prehistory and the beginning of the development of writing. Drawing is the construction of mental images, it is the representation of the student's knowledge, memory, and subsequent imagination. Through it it is also possible to identify the emotional state, feelings and frustrations that identify the individual characteristics of each student (Ehrlén, 2009; Al-Balushi at al, 2016). And it is for these reasons that it serves as a portrait of the individual's cognitive structure, providing the teacher with an analysis and assessment of the degree of abstraction and learning of the student in each content.

The drawing can be considered a universal language because a word written in different ways in several languages can be represented by a single design. Drawing is part of scientific development and has its maximum expression in Arts and Literature.

The act of drawing refers to our conception and vision of the world. When the student draws on the board what a word from a certain scientific area represents to him, he is expressing his knowledge about the subjects studied in the classroom, he exercises the right side of the brain (Edwards, 2005) contributing to the improvement of your memory, the development of your imagination, stimulates your creativity (Ainsworth et al, 2011), concentration and improves the systematization of ideas, which later can help you in improving your written and verbal language.

Regarding the development of verbal language, it is important to highlight a study of clinical analysis that shows the use of drawing and the act of drawing for the development of speech in deaf children (Araújo & Lacerda, 2008), such aspect shows the potentialities of the scope of the proposed use of the game Designing Science. The game can be used not only as a methodology to complement learning, social interaction, and evaluation, but also has a deep inclusive character and psychopedagogical development of the student. What we are looking for is to emphasize that the drawing exercise can bring benefits to students with or without hearing impairment, showing the countless advantages of using drawing for cognitive development and for clinical trials. In the next section we will look at the game in more detail.

## 2 Methodology and the rules of the game

2.1 Methodology

The game was built inspired by the popular game "Pictionary", with its structure and methodology focused on the Natural Sciences adapting, modifying, and adding the rules to the inclusive teaching modality, something that is not contemplated by the popular game mentioned above. The game board and cards were created following the new and unprecedented approach. To achieve the inclusion of deaf students, it was necessary to implement collective actions by all participants, replacing the speech of the meaning of the drawing made by the students on the blackboard with words written on small boards distributed individually to each team of players.

The game was applied, on average, to a total of 50 (fifty) students divided between two classes in state public high schools in the city of Parnaíba, in the state of Piauí and lasted two hours in each class. In one class there was no student with special needs and in the other there was only one deaf student, but a translator or sign interpreter was at his disposal.

The students of both classes, who were in the second year of high school, did not have much difficulty in understanding the dynamics of the game's operation, since they were being guided by three monitors of the Physics Degree course, in addition to the professor of Physics classroom.

2.2 Rules of the game

**Purpose of the game:** Make the team's own pawn be the first to cover the entire path of the board. Pawns are moved when players are able to make themselves understood through drawings, transmitting words and expressions from the sciences to teammates. The designer will have 60 seconds to draw the word or expression of the respective category.

There are six possible categories indicated by letters as shown in Table 2. The game is faster and more exciting if there are fewer teams and more players per team than the other way around. If there is an odd number of players, it doesn't matter that one team has one player more than the other.

| SCIENTIFIC GAME CATEGORIES | |
|---|---|
| **LETTERS** | SCIENTIFIC AREAS |
| A | ASTRONOMY AND GRAVITATION |
| B | BIOLOGY |
| E | ELECTROMAGNETISM, WAVES AND OPTICS |
| M | MECHANICS AND FLUIDS |
| C | CHEMISTRY |
| T | THERMODYNAMICS AND MODERN PHYSICS |

Table 2. Scientific categories of the game and their identification letters.

2.2.1 How to play

The starting house is a house **everyone play**. At the beginning of the game any participant can draw so that everyone can guess the meaning of the drawing. The team that gets it right will be entitled to the next move.

The designer cannot use physical or verbal communication, no matter how small. It is also not allowed to use letters or numbers. In the next round, instead of his team, the designer rolls the dice to find out which category of word will be drawn. Then take a new letter and draw the word of the drawn category. In Figure 1 we have the image of the game board. And in Figure 2, a game card is illustrated, where the different words associated with their respective areas of the natural sciences can be seen and the numbering corresponding to the faces of the dice, emphasizing that if the team gets the word right, that number will also be the quantity of squares that must be advanced in the game.

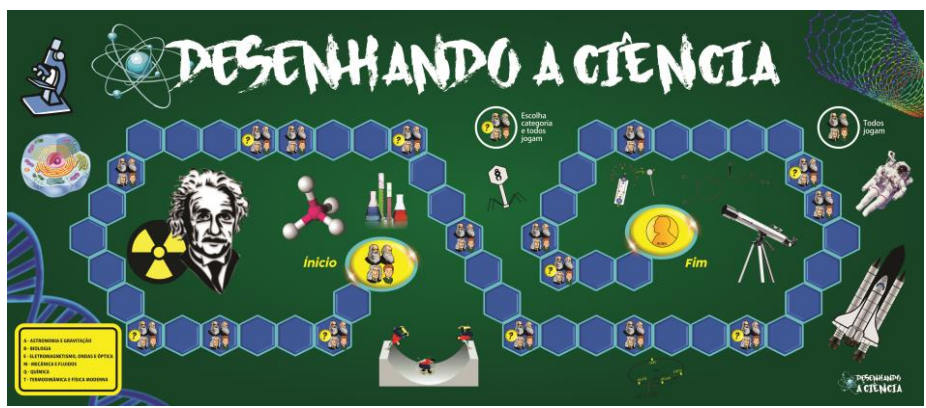

Figure 1. Game board template.

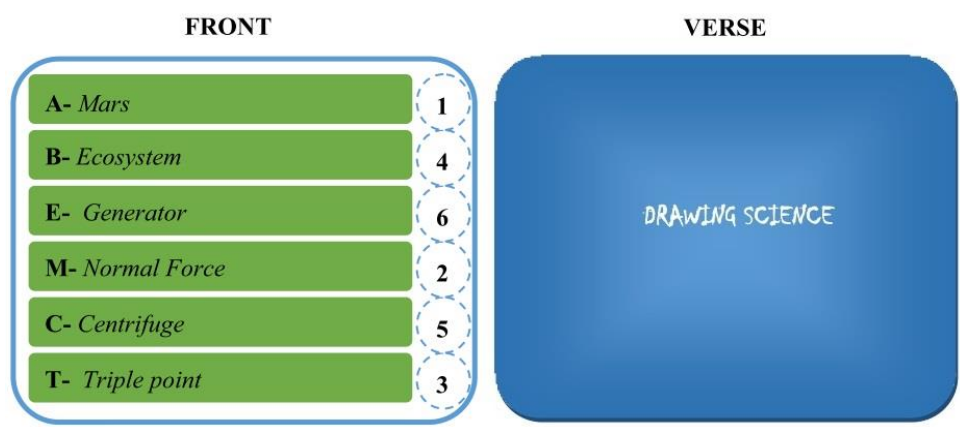

Figure 2. Card template with the words and their respective areas. The numbers on the right side, on the front of the card, indicate the number of squares that the teams must advance on the board if they hit the drawn word.

The board, after printing, has dimensions of 66 cm x 28.5 cm and the cards of 7.82 cm x 6.2 cm.

**Everyone play:** This house is represented by the images of four great scientists in the areas of Physics, Chemistry, Biology and Astronomy: Charles Darwin, Marie Curie, Isaac Newton and Galileo Galilei. When a team's pawn is in a square where **everyone play**, all teams participate in the round. The designer must roll the dice to know which category of the word he will draw. Then he takes a card, so that no one else sees it and starts to draw, and the team that gets it right first will walk the number of squares indicated on the card next to the correct word.

If two or more teams guess the word at the same time, a new card is drawn and a new round of **everyone play** is performed with only the teams that tied. If nobody gets the word right, neither team counts points, and the turn is passed to the next team.

**Choose category:** In some squares of the board the team at the time has the right to choose the category of the word that will be drawn, and everyone play. This square is similar to **everyone play** but comes with a question mark. In this case, there is no need to roll the dice, so the player will be able to choose the word that he considers the easiest, or the one that will give him the most points.

**Winner:** To win the game, a team must arrive with its pawn first on the last square represented by the medal awarded to the Nobel Prize winners by the Swedish Royal Academy of Sciences. It is not necessary to obtain the exact number of points to enter the last square on the board.

**Game-mime options:** Before the game starts, participants can choose to play by doing mimic instead of drawings. This decision must be accepted by all players. In this case, drawings will not be allowed during the game, only mimes! The rules for mimes are the same as for drawing.

2.2.2 Inclusive game mode

Once there are players with hearing impairment, the game allows integration, the rules remain the same, except:

**The game:** The designer will have 60 seconds to draw the word or expression of the respective category. However, your team members can try to guess the word when the designer finishes the drawing, or when 60 seconds elapse. The members of the group start to set the time again and organize themselves so that everyone can write a word in the white board and show the designer. After 60 seconds have elapsed and you don't get the word right, it is passed on to the next team.

**Everyone play:** The designer will have 60 seconds to draw the category word or expression. However, the members of his team or of the opposing teams only try to guess the word when the designer finishes the drawing, or when 60 seconds elapse. The timer starts again, and each team will have its board, and each team will be able to write a word after the drawing is finished. The team that shows the right word first, wins. After 60 seconds have elapsed and they haven't got the word right, no one receives points, and the turn is passed on to the next team.

Participants will not be able to try to correct by verbalizing words, only written words will be accepted. The game is suitable for high school students, undergraduate courses, and continuing education courses for teachers. The complete game in Portuguese and English versions is available for free and can be edited at the link: https://bit.ly/3bpPRYT.

**3  Results and discussions**

In the classes where the teaching and evaluation proposal was applied a good part of the contents that are covered in the game had already been seen in the subjects of Chemistry, Biology and Physics, which facilitated the performance of the students when drawing the words, emphasizing that they managed to do well even with the words associated with content seen in the previous series, in the previous series and with words in the area of Astronomy, which demonstrates how much the students had a consolidated cognitive structure and open to abstractions and the absorption of new knowledges. In figure 3 we have a photo of the game application.

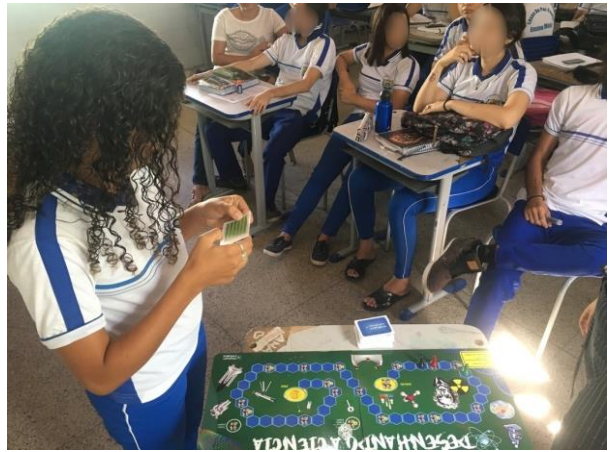

Figure 3. Students playing with game.

In contrast, when cards appeared with more advanced words in the deck associated with specific areas of Chemistry, Biology and Physics, they presented some difficulty, due to the fact that they did not know the subject, however, even in these situations, they used their imagination and made abstractions to draw something they didn't know, as can be seen in figure 4.

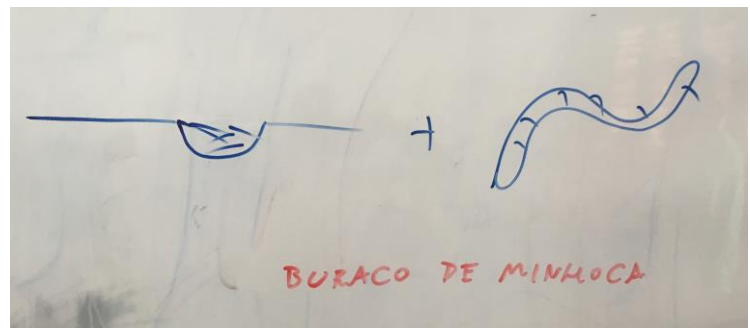

Figure 4. Student drew the word wormhole.

The word on the game card referred to the area of Astronomy and Gravitation. The Lorentzian wormhole is the term coined by physicist John Wheeler, in 1957 in Annals of Physics, to designate a singularity of space-time in the context of General Relativity, whose frontier is topologically trivial and which non-trivially connects two distinct points of space-time, see figure 5. They are also known as Schwarzschild wormholes or Einstein-Rosen bridges (Einstein & Rosen, 1935), being modeled after specific vacuum solutions of Einstein's field equations.

The student not only managed to draw the wormhole, but the members of his team were able to easily guess what it was about and got the word right. The theory of General Relativity geometrizes space, that is, it attributes the curvature of space-time to the causes of gravitational effects and in a simplified way the term wormhole is used to designate the Einstein-Rosen bridges, as this structure predicted by Relativity General can be seen as a tunnel that connects two distinct points of space-time, and in this figurative aspect the student's drawing is not wrong. In this situation it is good to emphasize the importance of the teacher, as a mediator of knowledge and so that he must intervene and provide the explanation and the correct context for the meaning of the word that was drawn. From the teacher's explanation, the student will have additional information and knowledge that will serve for the development of their cognitive structure.

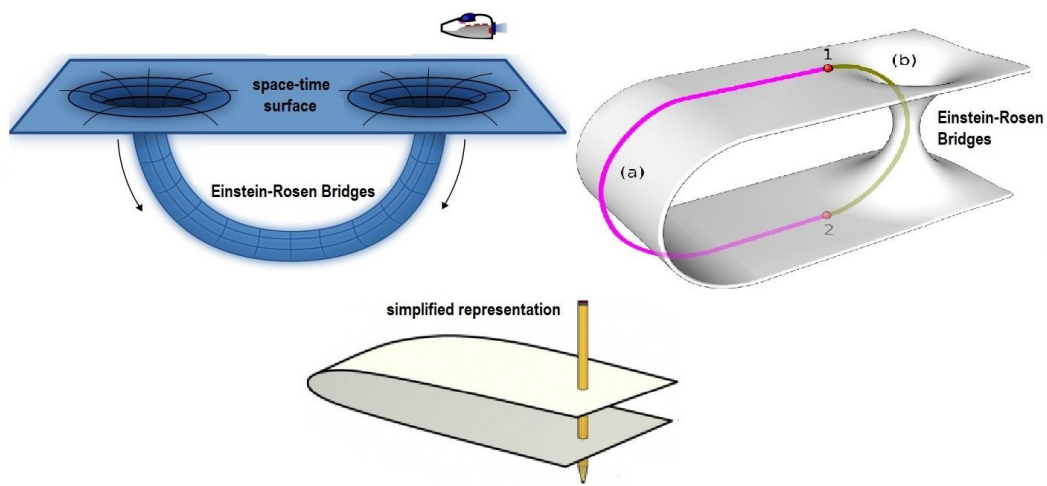

Figure 5. Top left: 2D analogy for a wormhole. Top right: Geometrical view of curved space-time (a) the magenta line: the longest way through normal space-time surface, (b) the green line: the short way through the wormhole. Down in the center: simplified representation of the curvature of space-time and the wormhole using paper and pencil. Source: Wikimedia Commons.

In the deck of cards there were words of different levels in all areas of the Natural Sciences, and whenever a player had difficulty drawing or the team misinterpreted the drawing, the game monitors and the room teacher explained the word and discussed its meaning. The card was placed back in the deck so that in another round, if the word occurred again, the student would be able to draw it and his team to interpret it, and within that aspect it also functioned as an assessment tool.

In figure 6 we have another interesting drawing in which another student managed to draw the word electric field. He could have used the usual diagrams seen in textbooks where the electric field is represented by means of force lines, only once again he used creativity and a very simple way to represent a subject that eventually his teammates did not have. seen in the classroom and still managed to get it right.

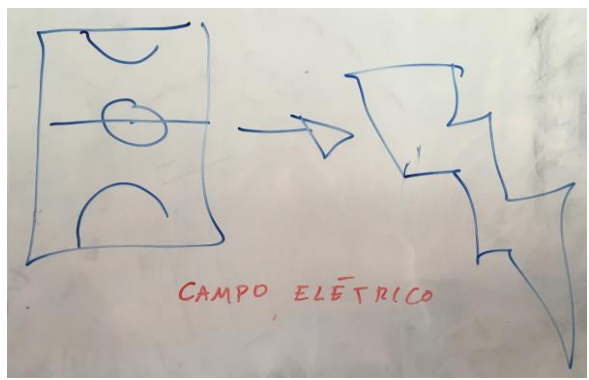

Figure 6. Student drew the word electric field.

The words written in red on the pictures were made by the monitors after the students got it right and the round was finished. It should also be noted that it is a race against time, it takes 60 seconds to give a correct answer, so the student looks for the simplest and most direct way to represent the word, abstracting information from other objects of his experience and knowledge, creating a correlation between them in order to have a figurative meaning.

In general, we can say that a field is a physical entity that has a real or complex existence, and may have a scalar, vector, and tensor nature; it permeates space-time, associating a physical magnitude to each point, allowing the interaction between particles. The fields, in their most abstract essence, are energy and, consequently, in high energies and relativistic regimes, they are also matter; so, the gage bosons are the result of quantizing fundamental fields (quantizing means dividing into discrete parts).

The vector fields associate only one vector to each point in space, such as the electric (**E**), magnetic (**B**) and velocity fields in a fluid (**v**). These fields can be obtained by using scalar field gradients. In the drawing of figure 6 the student associated the spatial region of interaction with a soccer field, which also delimits a spatial region where people interact. To designate the type of interaction the student used the design of a lightning bolt (electrical discharge), combining the two drawings we have a logical representation of the electric field.

Below we have in figures 7 and 8 drawings from the biology area, where the student drew, and made himself understood by his colleagues, the words fertilization and cell wall, respectively.

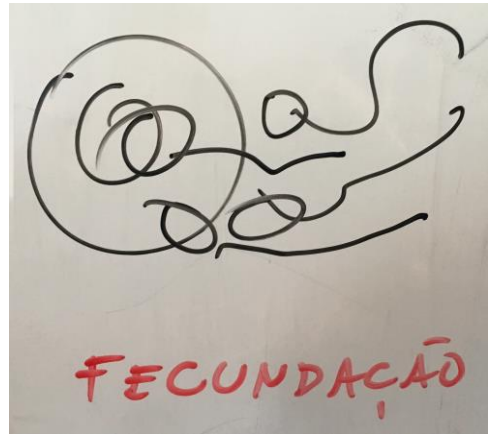

Figure 7. Student drew the word fertilization.

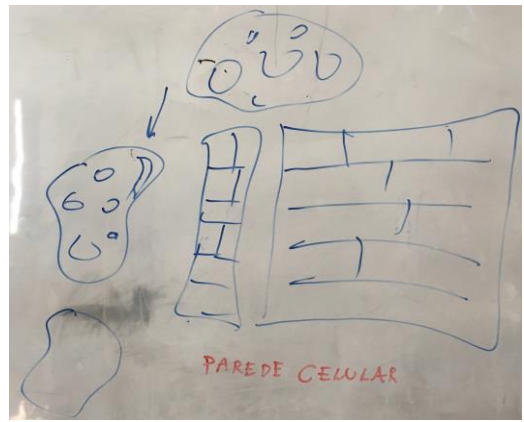

Figure 8. Student drew the word cell wall.

Only plant cells, bacteria and fungi have a cell wall, which surrounds the plasma membrane and has a structural function of maintaining cell shape and controlling its growth, in addition to preventing ruptures, protecting the cell from water ingress, and preventing cell death by Cytolysis (Sharp, 1921). In figure 8 the student represented the structural function of the cell wall by drawing a brick wall, and next to it a representation of the cell itself was drawn. Figure 9 shows the technical representation of a plant cell structure with all its components and makes a comparison with an animal cell and a bacterium. The cell wall in plants consists of organic substances such as cellulose, lignin, proteins and lipids, and inorganic substances such as crystals and silica. In bacteria it is usually composed of polysaccharide polymers linked to proteins such as murein (Alberts et al, 2002). Antibiotics such as penicillin act by inhibiting the bacterial cell wall synthesis process during binary division (Romaniuk & Cegelski, 2015).

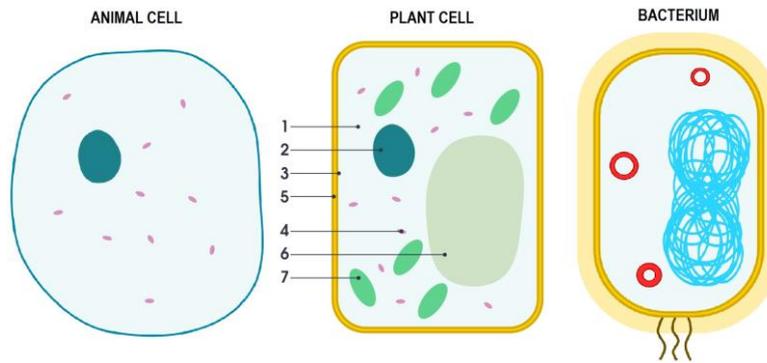

Figure 9. Comparative image of animal (without cell wall), plant, and bacterial cells. In the center we have a simple diagram of a plant leaf cell, with its constituents numbered for identification: 1- Cytoplasm, 2- Nucleus, 3- Cell membrane, 4- Mitochondrion, 5- Cell wall, 6- Permanent vacuole and 7- Chloroplast. Source: Wikimedia Commons.

In figure 10 we have a drawing of the Chemistry area, where the student drew, and made himself understood by his colleagues, the word coal. It is noted that he highlighted the charcoal bag and then the combustion process, which refers to charcoal, obtained by carbonization of wood (carbonization is a chemical reaction that removes hydrogen and oxygen from solids, so the remaining matter is mostly carbon) unlike mineral coal obtained in strata called coal layers being constituted by carbon, sulfur, hydrogen, oxygen, and nitrogen.

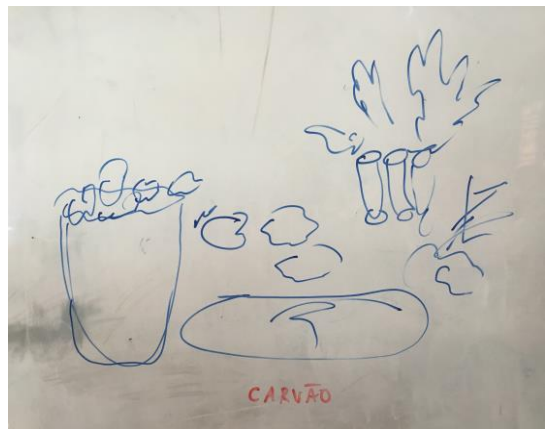

Figure 10. Student drew the word coal.

In figure 11 we have a drawing of the area of Thermodynamics in which the student drew one of the states of matter, a solid, and to make himself understood, he still represented the phase transition processes, the transition stages are clearly noted by the drawing from liquid to solid (solidification) and from solid to gaseous (sublimation).

Realize by these results that the game was successful in what it proposes that it is to complement the learning in a playful, fun way and to evaluate the cognitive structure of the students to investigate the degree of abstraction of scientific knowledge focused on the Natural Sciences.

The game was successful in the two classes in which it was applied, that is, both the class without a special student and the class that had a deaf student presented satisfactory performance and without difficulty in interaction and socialization. Highlighting that the presence of a sign language translator in one of the classes was of great importance for the initial presentation of the rules to the hearing impaired student, after that the game developed naturally.

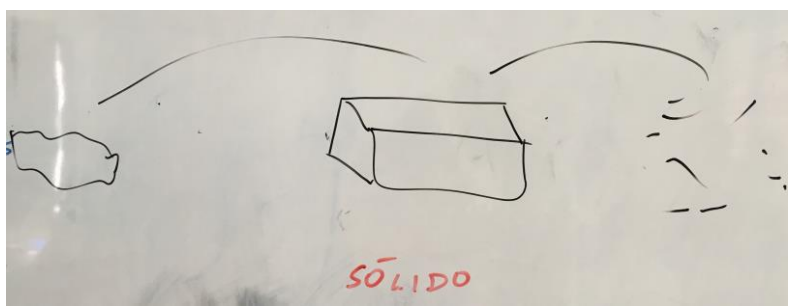

Figure 11. Student drew the word solid.

One of the great challenges of the application of the game was the participation of the deaf student. We realized that with the commitment of all students to follow the rules of the game in inclusive mode, using the small tables to write the word drawn on the blackboard, the game went on normally, with the same dynamics and the deaf student was very comfortable and even made the best drawings, as can be seen in figure 12, that shows the student with hearing impairment drawing the word telescope. It was a great joy for everyone to see how much he got involved, interacted, and had fun with the game and with his classmates.

With the possibility of playing in the modality using mimics, we opened a range of possibilities for the inclusion of deaf people, since the players must use sign language to communicate with the members of their team. At this point, who knows, the game could be used to teach sign language in undergraduate courses for teacher training.

The successful application of the game can be attributed to its playful and intuitive character, it has words related to science fiction movies, TV series, Robotics, space exploration, Engineering, renewable energy, and Medicine, promoting interdisciplinarity with the integration of different areas in different contexts, creating a parallel between scientific knowledge and common sense, instigating curiosity, and the fascination of students.

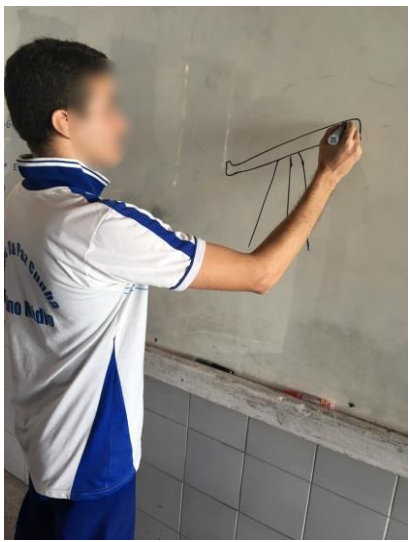

Figure 12. Hearing impaired student drawing the word telescope.

The students reported that they had a lot of fun playing, the dynamics of the game favored teamwork, inclusion and discussions focused on interpretation. They said that the game helped to consolidate the learning in the classroom and at the same time made possible the acquisition of new knowledge and the contact with new scientific areas. The components of each team were able to understand the abstraction of the designer by sharing his worldview, which led to the assimilation of information and the development of the imagery language.

As the game will be made available for free in an editable version, we leave as suggestions to educators that they can implement improvements in the game, making adjustments to their cultural and educational reality. For example, the areas we use in the charts: Astronomy and Gravitation; Biology; Electromagnetism, Waves and Optics; Mechanics and Fluids; Chemistry; Thermodynamics and Modern Physics, were chosen within the context of a physics teacher training course, so we gave more emphasis to the words in this area. However, we think it would be very interesting to make changes in the future and use the six areas defined by the National Curriculum Parameters (PCN): Movements: variations and conservation; Heat, environment and energy uses; Sound, image, and information; Electrical and telecommunications equipment; Matter and radiation; Universe, Earth, and life. This choice of areas makes the game even broader. Emphasizing that even choosing more physics words, the game includes scientific, technological, environmental and energy generation applications.

**4  Final considerations**

The teaching-learning process based on the learner's cognitive development must be aimed at building a dynamic and creative mentality in the student. In view of this aspect, this work explored the possibility of using a board game entitled Drawing Science as a tool to complement learning, assessment, scientific dissemination, interdisciplinarity and school inclusion in high school, also aiming at the formation of physics and teaching teachers Sciences. It was investigated if the students' learning about the contents studied in the classroom occurred in a significant way.

Based on what was exposed in this article, it can be seen, from the analysis of the drawings and observations made during the application of the game that students were able to develop and consolidate their cognitive structure, since they made abstractions and associations between knowledge previous knowledge of the world on general science topics, with more complex themes and focused on scientific research. They created well organized and illustrative drawings, exercising the right brain, contributing to the improvement of their memory, the development of their imagination, stimulation of their creativity, concentration, and improvement in the systematization of ideas, which can later help them in the improvement of your written and verbal language.

The proposal of the game as an inclusive methodology was successful since the hearing-impaired student was able to express himself and abstract the mental image of the words in the game and made himself easily understood by all the other members of his team.

The scope of the game is very wide, it can be easily translated into the native language of the user, it can be adapted to any culture, educational reality, and level of education. It can be played anywhere, including at home, where the student can play with his family.

For future teachers, we hope that the experience will serve to improve the pedagogical practice, to encourage reflections on teaching methodologies, to see the Sciences not only as a subject quantified by means of mathematical formulas and equations but like in the tale Le Petit Prince by Antoine de Saint-Exupéry, it can be understood through drawings, games and practical applications.

**Acknowledgment**

The authors would like to thank the students of the Physics Degree course Sandro Emerson Gomes, Gabriel Correia, Heully Lima, João Batista Pereira and Alberto Senna who contributed directly and indirectly to the realization of this work. And we especially thank Matheus Florêncio Fernandes for revising the text.

.